\documentstyle[12pt]{article}

\begin{document}
{\sf \begin{center} \noindent {\Large \bf Stretch-Twist-Fold and slow filamentary dynamos in liquid sodium Madison Dynamo Experiment}\\[3mm]

by \\[0.3cm]

{\sl L.C. Garcia de Andrade}\\

\vspace{0.5cm} Departamento de F\'{\i}sica
Te\'orica -- IF -- Universidade do Estado do Rio de Janeiro-UERJ\\[-3mm]
Rua S\~ao Francisco Xavier, 524\\[-3mm]
Cep 20550-003, Maracan\~a, Rio de Janeiro, RJ, Brasil\\[-3mm]
Electronic mail address: garcia@dft.if.uerj.br\\[-3mm]
\vspace{2cm} {\bf Abstract}
\end{center}
Recently Ricca and Maggione [MHD (2008)] have presented a very
simple and interesting model of stretch-twist-fold dynamo in
diffusive media based on numerical simulations of Riemannian flux
tubes. In this paper we present a yet simpler way of analytically
obtaining fast and slow dynamo, generated by by the curvature energy
of magnetic filaments in diffusive media. geometrical model for the
galactic or accretion disk dynamo in shear flows is presented. In
the fast dynamo case it is shown that the absence of stretching
leads to the absence of fast dynamos and when torsion of filaments
vanishes the dynamo action cannot be support as well. This is the
Cowling-Zeldovich theorem for planar flows. Isotropy of the magnetic
fields hypothesis is used to compute the fast nature of dynamo. A
similar result using non-holonomic Frenet frame has been recently
obtained for filamentary dynamos [Garcia de Andrade, AN (2008)]. The
stretch-twist-fold (STF) filamented models discussed here may serve
to formulate future experiments in the Madison Dynamo Experiment
(MDE) facility [Nornberg et al, Phys Plasmas \textbf{13} (2008)].
Though their running experiments deal with magnetic fields that are
orthogonal to the flow, other topologies are also used here, with an
eye in future experiments. Unstretched filaments in MDE shows Vishik
anti-fast dynamo action applies [Garcia de Andrade[Phys Plasmas
\textbf{15} (2008)].\vspace{0.5cm}

\newpage
\section{Introduction} Earlier Chui and Moffatt \cite{1} have investigated the role of Riemannian metric
and its back reaction on the helicity of magnetic twisted flux
tubes. More recently the effects of the Riemann metric on the
stretching and dynamo action have been also investigated \cite{2,3}
in plasma dynamo flows \cite{4}. The role of stretching is known to
be of utmost importance for dynamo action \cite{5}. Presence of
diffusion is actually fundamental in the Vainshtein-Zeldovich
stretch-twist-fold (STF) dynamo mechanism \cite{6} for folding and
merging processes in flux tubes, result that has been shown by Ricca
and Maggione \cite{7}. STF has been recently demonstrated
experimentally by Carey Foster and the Madison Dynamo Experiment
(MDE) group \cite{8}. Thus this fast dynamo mechanism is one of the
most well tested methods of obtaining fast dynamos. Contrarily to
what appears, the existence and discovered of fast dynamos are not
so easy due to the fact that simplicity criteria are in general
related with the symmetry processes that in general, due to
Cowling's and Zeldovich's anti-dynamo theorems, joint with diffusion
destroys dynamo action enhanced by advection. Presence of Frenet
holonomic frame turns problem geometry still more simple, than the
other techniques, however, not so simple as the Riemannian flux
tubes. By the way this seems to be the main reason, that the
Riemannian plasma kinematic dynamo solutions of the self-induction
equation \cite{2,3}, have ended with slow rather than fast dynamos.
In this paper, a throughly investigation of the kinematic dynamo
problem in the holonomic (depending only of the coordinate along the
magnetic filaments), shows that a fast dynamo can be obtained where
the magnetic Reynolds number is bounded from below by the inverse of
the constant curvature, while slow dynamos can be obtained by
another bound based on the total kinematic curvature energy. The
growth rate is computed in the fast dynamo case but mainly our
conclusions including also the slow dynamo case, are based on the
magnetic energy of the dynamos, obtained from the self-induction
equation. The slow dynamo is obtained by the limit of the vanishing
of diffusion limit and the growth rate ${\gamma}>0$. Actually the
fact of the matter is that, the filament investigation can also
produce a better understanding of the magnetic axis behaviour of
twisted magnetic flux tubes, where the role played by the twist is
here, played by torsion, of course having in mind that, twist may
have a torsion contribution. For example, a straight flux tube
cannot have a torsioned magnetic geometrical axis. This paper is
organised as follows: Section 2 addresses a brief review of Frenet
frame equations, useful for the reader to be able to follow the rest
of the paper. Section 3 presents the investigation of the absence of
stretching by only constraining the advection-stretching term to
vanish. Section 4 discusses the fast dynamo and slow cases in terms
of the magnetic energy and its relation with curvature kinetic
energy integral. Section 5 presents the orthogonal MDE case beteen
magnetic field and flows in diffusive media. Conclusions are
presented in section 6.\newpage
\section{Dynamo filamented flows in holonomic Frenet
frame} This section presents a brief review of the Serret-Frenet
holonomic frame \cite{9} equations that are specially useful in the
investigation of fast and slow filamented dynamos in
magnetohydrodynamics (MHD) endowed with magnetic diffusion. Here,
the Frenet frame is attached along the magnetic flux tube axis,
which possesses Frenet torsion and curvature \cite{11}. These
mathematical objects completely topologically determine the
filaments. Besides, one needs some dynamical relations from vector
analysis and differential geometry of curves, such as the Frenet
frame $(\textbf{t},\textbf{n},\textbf{b})$ equations
\begin{equation}
\textbf{t}'=\kappa\textbf{n} \label{1}
\end{equation}
\begin{equation}
\textbf{n}'=-\kappa\textbf{t}+ {\tau}\textbf{b} \label{2}
\end{equation}
\begin{equation}
\textbf{b}'=-{\tau}\textbf{n} \label{3}
\end{equation}
The holonomic dynamics considered here considers a steady evolution
of the tubes, which implies that the Frenet frame
$(\textbf{t},\textbf{n},\textbf{b})$, legs are vanish with respect
to time derivative. As one shall see in the next section this
considerably simplifies our task of finding solutions of
self-induction equations, avoiding some of the problems introduces
by curvature and torsion of the magnetic field topology and the flux
tube axis. Frenet frame with the help of equations
\begin{equation}
{\partial}_{t}{\textbf{t}}=[{\kappa}'\textbf{b}-{\kappa}{\tau}\textbf{n}]
\label{4}
\end{equation}
\begin{equation}
{\partial}_{t}{\textbf{n}}={\kappa}\tau\textbf{t} \label{5}
\end{equation}
\begin{equation}
{\partial}_{t}{\textbf{b}}=-{\kappa}' \textbf{t} \label{6}
\end{equation}
Together with the flow derivative
\begin{equation}
\dot{\textbf{t}}={\partial}_{t}\textbf{t}+(\vec{v}.{\nabla})\textbf{t}
\label{7}
\end{equation}
From these equations one is able in the next sections to write down
the expressions for the solenoidal magnetic and filamented flows
which allows us to split the self-induced magnetic equation generic
flow
\begin{equation}
{\nabla}.\textbf{v}=0\label{8}
\end{equation}
\begin{equation}
{\nabla}.\textbf{B}=0\label{9}
\end{equation}

\begin{equation}
\textbf{B}=B_{s}(r)\textbf{t}+B_{\theta}(r,s)\textbf{e}_{\theta}\label{8}
\end{equation}
In the section $3$ one shall solve the diffusion equation in the
steady case in the holonomic Frenet frame as
\begin{equation}
{\partial}_{t}\textbf{B}={\nabla}{\times}(\textbf{v}{\times}\textbf{B})+{\eta}{\Delta}\textbf{B}
\label{10}
\end{equation}
Therefore in the next section one shall address the stretching term
$(\textbf{B}.{\nabla})\textbf{v}$, which is present in equation
(\ref{10}) due to the vector analysis identity
\begin{equation}
{\nabla}{\times}(\textbf{v}{\times}\textbf{B})=-(\textbf{v}.{\nabla})\textbf{B}+(\textbf{B}.{\nabla})\textbf{v}
\label{11}
\end{equation}
The first term is advection and the second is the stretching term.

\newpage
\section{Unstretching magnetic filaments} In this section we
shall consider the stretching part of the self-induction equation
\begin{equation}
{d}_{t}\textbf{B}=(\textbf{B}.\nabla)\textbf{v}+{\eta}{\nabla}^{2}\textbf{B}
\label{12}
\end{equation}
By assuming, in the first example, that the filamentary stationary
flow is given by
\begin{equation}
 \textbf{v}(s)=v_{s}\textbf{t}+v_{n}\textbf{n}+v_{b}\textbf{b} \label{13}
\end{equation}
while the magnetic field is purely toroidal as $\textbf{B}$ as
\begin{equation}
\textbf{B}(t,s)=B_{s}(t)\textbf{t} \label{14}
\end{equation}
The only time dependence of $B_{s}$ comes from the solenoidal
property of the magnetic field above. Thus the stretching of the
filamented flow is given by
\begin{equation}
(\textbf{B}.\nabla)\textbf{v}=B_{s}{\partial}_{s}\textbf{v}=B_{s}[({v'}_{s}-{\kappa}v_{n})\textbf{t}+({v'}_{n}+{\kappa}v_{s}+
{\tau}v_{b})\textbf{n}({v'}_{b}-{\tau}v_{n})\textbf{b}]\label{15}
\end{equation}
The first expression on the RHS of this last expression vanished due
to the incompressibility condition
\begin{equation}
{\nabla}.\textbf{v}= {v'}_{s}-{\kappa}v_{n}=0 \label{16}
\end{equation}
By assuming that the flow is planar, to test Cowling-Zeldovich
theorem, $({\tau}=0)$ the stretching condition yields the two simple
ODEs
\begin{equation}
 {v'}_{s}-{\kappa}v_{n}=0\label{17}
\end{equation}
\begin{equation}
{v'}_{n}+{\kappa}v_{s}=0 \label{18}
\end{equation}
Multiplying the equation (\ref{12}) by $v_{s}$ and the equation
(\ref{13}) by $v_{n}$ and subtracting the resulting equation
\begin{equation}
T={v^{2}}_{s}+{v^{2}}_{n}=c_{0} \label{19}
\end{equation}
where $c_{0}$ is an integration constant. This shows that the
kinetic energy is constant, which shows that the absence of
stretching cannot yield a growing kinetic flow energy capable to
induce dynamo action. To simplify the above computations one assumes
that the filament is helical, where the constant torsion
${\tau}_{0}$ equals the constant curvature ${\kappa}_{0}$. Thus
Cowling theorem of non-dynamo action for planar flows is valid. The
next example is much simpler. This case uses the following fields
\begin{equation}
\textbf{B}=B_{s}\textbf{t}+B_{n}\textbf{n} \label{20}
\end{equation}
\begin{equation}
\textbf{v}=v_{s}\textbf{t} \label{21}
\end{equation}
Due to the solenoidal (incompressibility) of the flow ${v_{s}}'$
vanishes and therefore the stretching term becomes
\begin{equation}
(\textbf{B}.\nabla)\textbf{v}={\kappa}_{0}v_{s}B_{n} \label{22}
\end{equation}
Now imposing the vanishing of stretching yields either that
curvature vanishes or normal component vanishes or yet that the
toroidal flow vanishes, each relation of these yields a weakness of
dynamo action.
\section{Fast and slow filamented dynamo flow}
In this section one shall consider the two examples considered above
in full detail and solve the self-induction equation in the presence
of stretching and diffusion to check for the existence of fast or
slow kinematical dynamos here the action of back reaction of Lorentz
flow is neglected. Let us now consider first the slow dynamo case
which is present in the first example of the last section. Since the
stretching terms were computed in the previous section, here one
shall only compute the diffusive and growth rate of the magnetic
fields as
\begin{equation}
{d}_{t}\textbf{B}=(d_{t}{B}_{s})\textbf{t}+{B}_{s}d_{t}\textbf{t}
\label{23}
\end{equation}
and
\begin{equation}
{\nabla}^{2}\textbf{B}=B_{s}[{\kappa}(-{\kappa}\textbf{t}+{\tau}\textbf{n}]
\label{24}
\end{equation}
Collecting these terms one obtains from the scalar equation along
the $\textbf{t}$ direction
\begin{equation}
\frac{1}{2}d_{t}(\int{{B_{s}}^{2}ds})=-{\eta}\int{{B_{s}}^{2}{\kappa}^{2}ds}
\label{25}
\end{equation}
Since due to the solenoidal character of a monopole free magnetic
field, the magnetic field $B_{s}$ toroidal component just depends on
the time coordinate, which reduces the last equation to
\begin{equation}
\frac{1}{2}d_{t}(\int{{B_{s}}^{2}ds})=-{\eta}{B_{s}}^{2}\int{{\kappa}^{2}ds}
\label{26}
\end{equation}
This expression shows clearly that the magnetic energy is
proportional to the curvature kinematic energy
\begin{equation}
K:=\int{{\kappa}^{2}ds} \label{27}
\end{equation}
Note however that when the diffusion constant, vanishes, the
toroidal magnetic energy is constant and therefore the dynamo is
toroidally slow. The other expressions for the energy can be
obtained for the other directions and we are able to say that the
dynamo is actually slow. Combining these three equations one obtains
the dynamo relation
\begin{equation}
\frac{1}{2}d_{t}(\int{{B_{s}}^{2}ds})=v_{s}{\kappa}_{0}\int{{B_{s}}^{2}ds}-\frac{{\eta}}{2}B^{2}
\label{28}
\end{equation}
In the limit of ${\eta}\rightarrow{0}$ from this expression, is easy
to see that the magnetic energy may grow under appropriated
constraints. Thus the fast dynamo action is possible under the STF
method exactly in the MDE. In the next section a more detailed
explicity example shall be done without the assumption of isotropy
$B_{s}=B_{n}$ used here to simplify computations.
\section{STF-MDE filament experiment}
In this section, we have considered the MDE to test STF, providing a
filamented model analogous to the previous ones, with only the
difference that the magnetic and velocity fields are orthogonal
$(\textbf{B}.\textbf{v}=0)$, which is used in the initial stages of
the MDE. The choice given is
\begin{equation}
\textbf{B}=B_{s}\textbf{t}+B_{n}\textbf{n} \label{29}
\end{equation}
and for the flow velocity
\begin{equation}
\textbf{v}=v_{b}\textbf{b} \label{30}
\end{equation}
Note that the advective term above does not vanish, since the cross
product $\textbf{v}{\times}\textbf{B}$ does not vanish. This yields
the possibility of the fast dynamo action as obtained in the Madison
experiment. The stretching term is given by
\begin{equation}
(\textbf{B}.\nabla)\textbf{v}=B_{s}[{v'}_{b}\textbf{b}-{\kappa}_{0}v_{b}\textbf{n}]
\label{31}
\end{equation}
In the unstretched case, where this expression vanishes,and since
the curvature ${\kappa}_{0}$ is assumed constant and non-vanishing
the only solution is the vanishing of $v_{b}$ and since, this is the
only flow component, the magnetic filamented flow is static which
does not support any dynamo action. This is Vishik´s anti-fast
dynamo theorem. Since curvature is the Frenet curvature of the
filaments, and not curvature of liquid metal device, the curvature
${\kappa}(s)$ may be non-constant. This yields the following
equation
\begin{equation}
(\textbf{B}.\nabla)\textbf{v}=B_{s}[{v'}_{b}\textbf{b}-{\tau}v_{b}\textbf{n}]
\label{32}
\end{equation}
The vanishing of this expression implies that the $v_{b}$ is
constant and that the torsion ${\tau}$ vanishes, which forces the
filament flow to be planar and by Cowling's theorem forbides dynamo
action. Now let us consider the whole self-induction equation. With
the help of the solenoidal property of the magnetic field, one
obtains
\begin{equation}
{\nabla}.\textbf{B}={\partial}_{s}B_{s}-{\kappa}_{0}B_{n}=0\label{33}
\end{equation}
after a long computation one obtains the following magnetic energy
relation for the $B_{n}$ component
\begin{equation}
\frac{1}{2}\frac{d}{dt}\int{{B^{2}}_{n}ds}=-2{\eta}{{\kappa}_{0}}^{3}\int{{B_{n}}^{2}\frac{({\kappa}_{0}-v_{b})}{{v'}_{b}}ds}+
{\eta}\int{[{B"}_{n}B_{n}-(2{B_{n}}'B_{n}+{B_{n}}^{2}){{\kappa}_{0}}^{2}]ds}\label{34}
\end{equation}
This expression shows that if the toroidal component did vanish and
only normal component of the magnetic field survives the dynamo
would be slow and not fast. Actually the by considering the
remaining component of the self-induction equation along the
binormal direction $\textbf{b}$
\begin{equation}
\frac{B_{s}}{B_{n}}=-2{\eta}\frac{{{\kappa}_{0}}^{2}}{{v'}_{b}}\label{36}
\end{equation}
This equation together with the solenoidal character of
$\textbf{B}$, given by expression (\ref{33}) yields the following
expression for the toroidal magnetic energy growth rate
\begin{equation}
\frac{1}{2}\frac{d}{dt}\int{{B^{2}}_{s}ds}=\int{{B_{s}}^{2}(2{v'}_{b}+{v"}_{b})ds}\label{37}
\end{equation}
This shows that the toroidal magnetic part of the of the growth rate
of the energy does not depend on diffusion and thus the possibility
of the fast dynamo through the STF method is guaranteed, such as in
MDE experiment.
\section{Conclusions} In the case of STF filamentary MDE model,
the initial magnetic field in the laminar non-turbulent phase is
orthogonal to the flow. Note that expressions for the magnetic
filaments may be chosen with appropriate topology and boundary
conditions in order to guarantee, dynamo action in the presence of
magnetic resistivity MHD equations. In this paper we propose
alternative topology between the magnetic lines and the flow in
order to guarantee fast kinematic dynamo action through the STF
method. Slow dynamo is obtained in some situations with curvature
energy integral bounds while , as shown in the MDE experiment, the
initial orthogonal relation between magnetic and flow fields, seems
to grant the existence of fast dynamos in the STF mechanism. The
other examples treated here, may be useful for designing another
experiments that could possibily be running in Madison dynamo
experimental facility in near future.
\section{Acknowledgements} Thanks are due to D Sokoloff for helpful discussions on the subject of this paper. I also appreciate financial
supports from UERJ and CNPq.


\newpage
\begin{thebibliography}{9}
  \bibitem{1} Y Chui, K Moffatt, Proc Roy Soc London A (1995).
  \bibitem{2} L. C. Garcia de Andrade, Physics of Plasmas
  \textbf{15},122106. L.C. Garcia de Andrade,Non-holonomic dynamo filaments
   as Arnold´s map in Riemannian space, Astronomical notes (2008) in press.
  \bibitem{3} L. C. Garcia de Andrade, Physics of Plasmas \textbf{14}, 102902 (2007).
  \bibitem{4} Z. Wang, V. Pariev,C Barnes, D Barnes, Phys Plasmas
  \textbf{9} 1491 (2002).
  \bibitem{5} S. Childress and A.D. Gilbert, Stretch, Twist and
  and fold, Springer (1993).
  \bibitem{6} S. I. Vainshtein, Ya B Zeldovich, Sov Phys Usp 15
  ,159 (1972). Ya B. Zeldovich, A Ruzmaikin and D Sokoloff, Magnetic
  fields in Astrophysics (1983) Gordon and Breach Publishers. V Arnold, Ya B Zeldovich, A Ruzmaikin and D D
  Sokoloff, Soviet Physics JETP (1981).
  \bibitem{7} R Ricca, Maggione, A Simple model for Stretch-Twist and Fold Dynamo, Talk delivered at MHD 2008-Nice.
  \bibitem{8} M Nornberg,E Spence, R Kendrick, C Jacobson and C B Forest, Phys
   Plasmas \textbf{13} (2006) 055901.
  \bibitem{9} P K Newton, The N-Vortex Problem: Analytical
  Techniques, (2001) Springer,145.
  \end{thebibliography}
  \end{document}